\documentclass[doublecol]{epl2}

\usepackage{graphicx,epsfig}

\title{Heterogenous Human Dynamics in Intra and Inter-day Time Scale}
\shorttitle{Heterogenous Human Dynamics in Intra and Inter-day Time Scale} 

\author{Peng Wang\inst{1,2} \thanks{\email{wangpenge@gmail.com}}\and Ting Lei\inst{2} \and Chi Ho Yeung\inst{2} \and Bing-hong Wang\inst{1,3} \thanks{\email{bhwang@ustc.edu.cn}}}
\shortauthor{P. Wang \etal}

\institute{
  \inst{1} Department of Modern Physics, University of Science and
Technology of China, Hefei 230026, China\\
  \inst{2} Department of
Physics, University of Fribourg, Fribourg, Switzerland\\
  \inst{3} The Research
Center for Complex System Science, University of Shanghai for
Science and Technology and Shanghai Academy of System Science,
Shanghai, 200093 China}


\abstract{In this paper, we study two large data sets containing the
information of two different human behaviors: blog-posting and
wiki-revising. In both cases, the interevent time distributions
decay as power-laws at both individual and population level.
As different from previous studies, we put emphasis on time scales and
obtain heterogeneous decay exponents in intra- and inter-day range
for the same dataset. Moreover, we observe opposite trend of
exponents in relation to individual $Activity$. Further
investigations show that the presence of intra-day activities mask
the correlation between consecutive inter-day activities and lead to
an underestimate of $Memory$, which explain the contradicting
results in recent empirical studies. Removal of data in intra-day
range reveals the high values of  $Memory$ and lead us to convergent
results between wiki-revising and blog-posting. }

\begin{document}

\maketitle

\section{Introduction}
\label{S1:Intr}

Thanks to the development of the information technology,
comprehensive data available from the internet give us valuable
insights into the pattern of human behaviors. Many recent studies of
human behavior focus on the distributions of inter-event time or
waiting time and report a heavy-tail both at the individual and population level.
Examples of empirical studies including
communication patterns of electronic mails\cite{dm7,dm8,dm1,dm2} and
surface mail\cite{dlett1,dm2}, web surfing\cite{dweb1,dweb2},short
message\cite{dmessage1}, movie rating\cite{dmov1}.

In all the above systems, the observed distributions of interevent
time goes as $\tau^{\alpha}$ with exponents ranging from 1 to 3.
Various mechanisms were suggested to explain the underlying
dynamics. One main class of mechanism is the priority-queue
model\cite{dm7,dm8}, which yields power-law waiting-time
distributions $p(\tau)=\tau^{a}$ with universal exponents a=1.0 and
1.5. Other mechanisms include the adaptive interesting
model\cite{m2}, the memory model\cite{m4} and the interaction
model\cite{m3}. A crucial assumption of all these models and
empirical studies is that the mechanisms driving human behaviors are
identical in all time scales. According to this assumption,
interevent time with length in minutes and in days are generated by
the same mechanism and follow the same scaling law. Even in the
cascading nonhomogeneous Poisson process\cite{dm1,dm2} which
emphasizes external factors such as circadian and weekly cycles,
the distributions still follow power-laws with identical exponent
over the whole range.

\begin{table*}
\begin{center}
\begin{tabular}{|l||c||c||c|}
\hline
 Human activity&Unit&Range&Exponent\\
 \hline \hline
\verb"Email"\cite{dm8,m4}&sec&intra-day&$1\ast$, 0.9\\
\verb"Correspondence"\cite{m4,dco1}&day&inter-day&$2.37\bigtriangleup$, $2.1\bigtriangleup$\\
\verb"Library loans"\cite{dm8}&min&intra-day&$1\ast$\\
\verb"Printing behavior"\cite{dprint1}&sec&intra-day&$1.3\bigtriangleup$\\
\verb"Visits of a web portal"\cite{dm8}&sec&intra-day&$1\ast$\\
\verb"Visits to the same URL"\cite{dweb1}&sec&intra-day&1\\
\verb"Visits to any page"\cite{dweb1}&sec&intra-day&1.25\\
\verb"Queries on AOL"\cite{dweb2}&hour&inter-day&1.9\\
\verb"Message on Ebay"\cite{dweb2}&hour&inter-day&1.9\\
\verb"Logging actions on Wikipedia"\cite{dweb2}&hour&inter-day&1.2\\
\verb"Movie rating"\cite{dmov1}&day&inter-day&2.08\\
\hline
\end{tabular}
\end{center}

\caption{Comparison of the exponents from different human
activities. The unit of interevent times and the time range in these
studies are shown in the table. $\ast$ corresponds to the
average of exponents from individual distributions;$\bigtriangleup$ corresponds to
the exponent from a single user; others are the ones from global
distribution} \label{tableWealth}
\end{table*}

Table 1 shows a collection of recent empirical results, including
the exponents and the unit of interevent
time and the time range where the power-laws were observed. In this
table, we simply classify the results into intra- and inter-day
behaviors. As we can see, for those data with unit in second or
minute, the studies are often focused on the intra-day interevent
time distribution; for those with unit in hour or day, only the
inter-day range was studied. None of these studies investigated both
the intra and inter-day behavior, though some noticed a hump in the
interevent time distribution caused by the circadian
rhythm\cite{dlinux1}. One case that had been studied intensively is
email and letter based communications, where some studies suggested that
mechanisms of the two activities are different, based on the
different exponents observed\cite{dm8}; others suggested that the
two are essentially the same based on the data collapse of
interevent time distributions\cite{dm2}. Limited attention has been
paid on the different time range in the two activities, as the
timestamp of email and letters communications are respectively in
the unit of second and day, and exponents are thus extracted from
different time range. By comparing the activities in the table, we
find that the inter-day exponents tend to be clearly higher than the
intra-day ones: the exponents of four of the five inter-day
activities are around or more than 2; all the exponents of the six
intra-day activities are around or a little more than 1.

It is, of course, insufficient to prove the above relationship only
by comparison between the exponents of different activities. We thus
aim to bring further evidence in this paper. Our work is based on
two data sets from different sources which record two kinds of human
activities: wiki page revising and blog posting\cite{dblog1}. The
heavy-tails are found in both intra- and inter-day part of the
distributions from these two activities. Our results show that even
for the same activity the exponents of these two ranges are
different.

Further evidences are obtained by examining the dependence of decay
exponent on individual $Activity$, the measure of how frequent the
action is taken. Zhou et al\cite{dmov1} found that the exponent
increases with $Activity$, which was further confirmed by
Radicchi\cite{dweb2}. It is noted that both analyses are conducted
in the inter-day range. In our case, we found the same dependence in
the inter-day range but remarkably a different behavior in the
intra-day range. It further demonstrates that the mechanisms
underlying intra- and inter-day human dynamics are different.

On the other hand, weak memory in human behaviors are observed in
system such as library loans and printing\cite{2}. However, other studies
show significant memory in some systems driven by
human\cite{dblog1,memory2,dprint1}. For wiki page revision, we found
seemingly weak memory. However, we observe a strong memory
comparable to that of blog-posting\cite{dblog1} by removal of
intra-day intervals and consider the inter-day ones only. It shows
that the memory of inter-day activities is underestimated as
intra-day activity mask the correlation between inter-day activities
in analysis. We suggest that it is the reason behind the apparent
weak memory in some human behaviors.

\section{Data sets description}

\subsection{Wikipedia}

Wikipedia (Wiki) is a free encyclopedia written in multiple languages
and collaboratively created by volunteers. Wiki contains millions of
articles which is produced by tens thousands of online volunteers.
When an article is revised by an user, a new version is created by
this user. The database we consider contains the timestamp and the
authors of all the revisions in the chinese Wiki. This data set is
composed of 9641842 revisions made by 81823 users between
26/10/2002 and 7/6/2009.

\begin{figure}[htb]
\centerline{\epsfig{figure=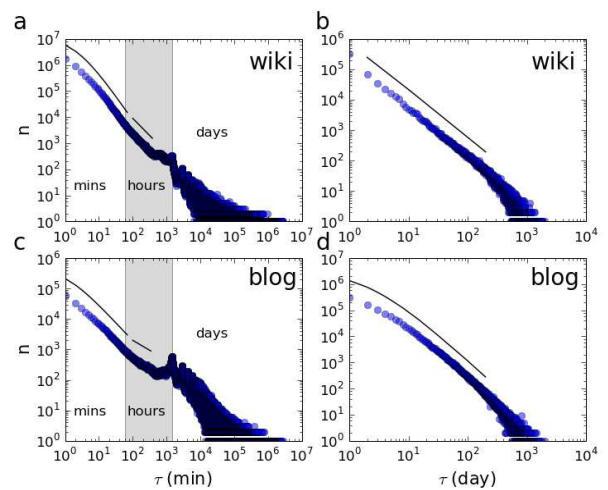, width=\linewidth}}
\caption{\label{Fig:RI:PDF}(Color online) The global distribution of
interevent time spanning the intra- and inter-day range. $n$ is
frequency. We fit the distributions with the ``shifted power-law":
$y\sim(x+a)^{-\beta}$\cite{book}. Figure (a) and (c) shows the
distributions of the intra-day range of wiki-revising and
blog-posting; Figure (b) and (d) shows the inter-day range. The
decay exponents are $\beta_{mins} \simeq 1.88$ and $\beta_{hours}
\simeq 1.32$ in (a), $\beta_{mins} \simeq 1.20$ and $\beta_{hours}
\simeq 0.66$ in (c); $\beta \simeq 1.57$ in (b), $\beta \simeq 2.02$
in (d).}
\end{figure}

\subsection{Blog}
Blog is a kind of so-called web2.0 applications emerging in recent
years, on which people post, read and comment articles from each
other\cite{blog1,blog2}. Our data was collected from the campus blog
website of Nanjing university(http://bbs.nju.edu.cn/blogall). Most
users are current or former students and teachers of Nanjing
university. As of 01/09/2009, there are 1627697 articles posted by
20379 users in this website. The first post is at 25/03/2003 when
the blog was established.

\section{Empirical analysis}

\subsection{The global distribution of interevent time in intra-day and inter-day range}

The timestamp of both the data sets is in precision of one minute.
Here, the interevent time $\tau$ is time interval between
consecutive actions, i.e. revising a wiki-page by the same user in
wiki or posting an article by the same user in blog. The global
distributions of $\tau$ for both data sets are shown in figure 1. As
we can see, the distributions can be divided into two parts: For the
intra-day range, the curves clearly show the heavy-tails; for the
inter-day range, they all show oscillation because of the circadian
periodicity that make it hard to observe the scaling law.

Even in the intra-day range, the power-law behavior is not
homogeneous in all time scale and a slight hump is observed at $\tau
\approx 100$ (see fig. 1(a) and (c) and fig 4 for clearer evidence).
We thus apply a piecewise fitting curve to show the change in
power-law exponents. For the range with $\tau < 100$ (within about 1
hour), the exponents of blogging and wiki-revising activities are
1.20 and 1.88; for the range with $\tau
> 100$(beyond 1 hour and within 1 day), lower values of 0.66 and 1.32 are found.

Figure 1(b) and (d) shows the distribution of inter-day interevent
time where a unit of one day is employed to eliminate the
oscillation. The heavy-tails in the inter-day range are shown
clearly in these two distribution. The exponent of blogging activity
is 2.02 which is significantly higher than the ones in intra-day
range, in agreement with the results obtained by comparing different
empirical studies in table 1. On the other hand, the intra-day
exponent of wiki-revising seem to be close to the inter-day one.
However, as we will see in following section, the empirical analysis
at group and individual levels demonstrate the different activity
pattern between the two ranges.

\begin{figure}[htb]
\centerline{\epsfig{figure=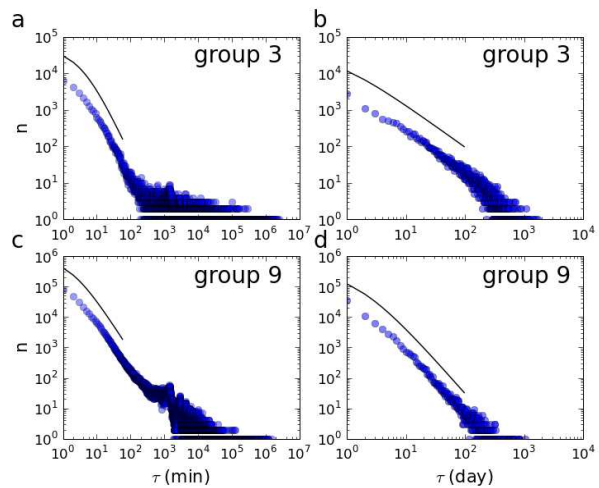, width=\linewidth}}
\caption{\label{Fig:RI:PDF}(Color online) The interevent time
distribution of wiki-revising at a group level (group 3 and group 9).
For distribution in the intra-day range, the range of fitting is
from 1 to 70. (a) and (c) correspond to the intra-day range and (b)
and (d) correspond to the inter-day range. The decay exponents are
$\beta \simeq 2.00$ in (a),$\beta \simeq 1.16$ in (b),$\beta \simeq
1.75$ in (c),$\beta \simeq 2.21$ in (d).}
\end{figure}

\subsection{Heterogeneous Dependence on Activity}

In this section, we will investigate further the features of intra-
and inter-day activity pattern. Firstly, we measure the average
$Activity$ $A_i$ of user i as $A_i=n_i/d_i$, where $n_i$ is the
total number of actions of user $i$ and $d_i$ is the time between
the first and the last actions. We then sort users in an ascending
order of $Activity$ and divide the entire population into 10 groups,
each of which have $M$ users ($M \approx N/10$ where $N$ is the total
number of users). The first $M$ users in the list belong to group 1,
and the last $M$ users are belong to group 10, etc. We only consider users
with $n_i,d_i>10$. For wiki, there are 14410 qualified users and
$M=1400$; for blog, there are 12827 qualified users and $M=1300$.
As different from previous studies\cite{dmov1,dweb2} which only focus
on the inter-day range, we investigate the dependence of the
exponent on $Activity$ in both the intra- and inter-day range. In
fig. 2, we plot the interevent time distribution of wiki for group 3
and 9 (which respectively correspond to average $Activity$
$\langle A \rangle=0.07, 1.12$). For the inter-day range, we get the
same dependence as the one obtained in other inter-day activities :
the exponents increase with $Activity$. Some exponents of inter-day
activities are small such as the one in logging action probably due
to the relatively low activity \cite{dweb2}. For the intra-day
range, this dependence is totally different: the exponents decrease
with increasing $Activity$ and the change is relatively smooth. In
fig. 3, we plot the exponent of the interevent time distribution of
wiki-revising and blog-posting as a function of $Activity$. Though
the values of exponents are different in these two cases, they show
the same features: the exponent and $Activity$ are positively
related in the inter-day part and negatively related in the
intra-day one.

\begin{figure}[htb]
\centerline{\epsfig{figure=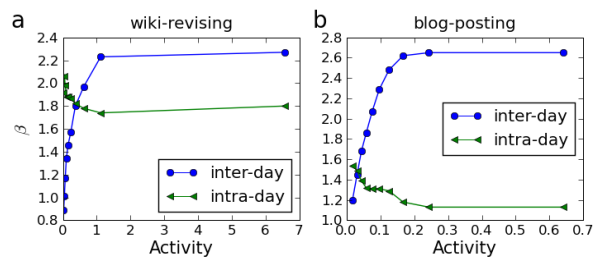, width=\linewidth}}
\caption{\label{Fig:RI:PDF}(Color online) Dependence of decay
exponents on $Activity$.}
\end{figure}

\subsection{Interevent time distribution for individuals}

To show further evidences for our conjecture, we look in detail the
behavior of individual agents. Figure 4 shows the cumulative
distribution of interevent time from four users, two are from the
data set of wiki and two are from the blog data set. An obvious
trend change is observed at $\tau \approx 1$ day. For the inter-day
range, all these distributions follow power-laws. The wiki users
often revise one page many times within a day but blog users seldom
post several articles in one day. Therefore, it is hard to study the
intra-day activity of blog-posting at the individual level as data
is insufficient in this range. For wiki-revising, the distributions
are even heterogenous within the intra-day range(see fig 4(a)),
which is consistent with the global one and shows further complexity
in the mechanism of human activity.

\begin{figure}[htb]
\centerline{\epsfig{figure=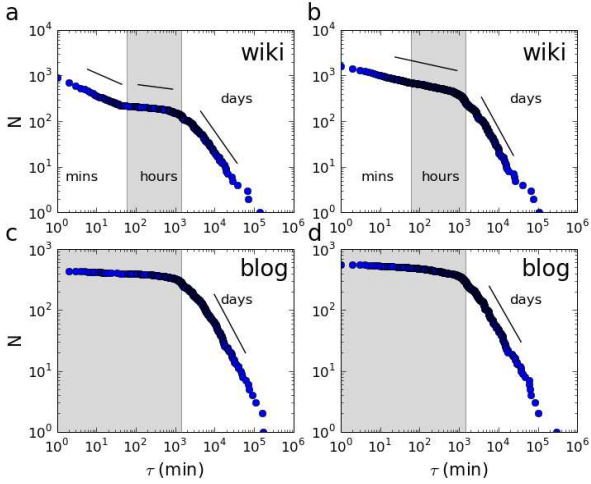, width=\linewidth}}
\caption{\label{Fig:RI:PDF}(Color online) The cumulative
distribution of interevent times of individuals. $N$ is the
cumulative frequency of intervals. User 1 and User 2 in (a) and (b)
are from wiki; User 3 and User 4 in (c) and (d) are from blog. The
decay exponents are $\beta_{mins} \simeq 0.38$, $\beta_{hours}
\simeq 0.11$ and $\beta_{days} \simeq 1.23$ in (a), $\beta_{hours}
\simeq 0.19$ and $\beta_{days} \simeq 1.57$ in (b); $\beta \simeq
1.22$ in (c), $\beta \simeq 1.13$ in (d)}
\end{figure}

\begin{figure}[htb]
\centerline{\epsfig{figure=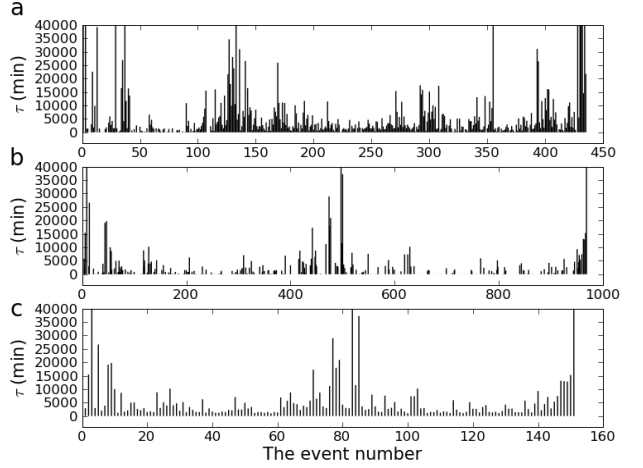, width=\linewidth}}
\caption{\label{Fig:RI:PDF}The interevent time of consecutive events
(a) User 3 in figure 4c. (b) User 1 in figure 4a. (c) User 1 after
deleting short interevent times which is less than 1000 mins.}
\end{figure}

\begin{figure}[htb]
\centerline{\epsfig{figure=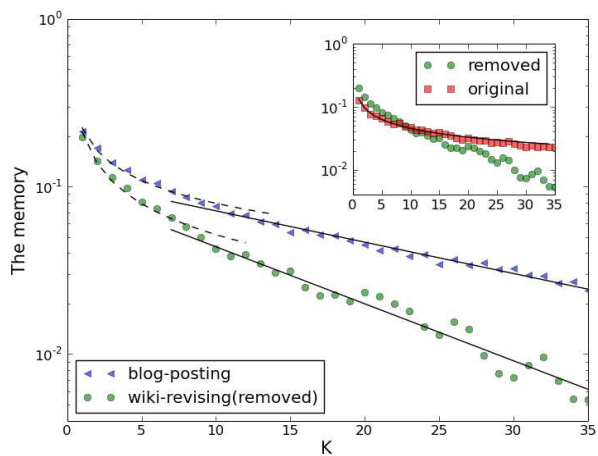, width=\linewidth}}
\caption{\label{Fig:RI:PDF}(Color online) The average $M_k$ of
all qualified users in blog-posting and wiki-revising after data
removal with different $k$. The comparison between the results before
and after data removal is shown in the inset. For the one of
blog-posting, $M_{k}$ decays as a power law when $k<10$:
$M_{k}=0.23*k^{-0.45}$; when $k>10$, it decays exponentially:
$M_{k}=0.1*e^{-k/23.22}$\cite{dblog1}.  For the original data, it
decays as a power law over whole range:$M_{k}=0.13*k^{-0.47}$. After
data removal, when $k<9$: $M_{k}=0.61*k^{-0.21}$; when $k>9$:
$M_{k}=0.10*e^{-k/12.76}$. To avoid characterizing users whose
number of actions is too small, we consider only the qualified users
of the two data sets and calculate the memory of all these users
with $k$ ranging from 1 to 35 (for wiki, a total of 809 users with
number of revisions more than 800 and frequency of long intervals
($>1000$ mins) more than 100 are considered; for blog, a total of
2126 users with more than 200 posts and frequency of long
intervals ($>1000$ mins) more than 200 are considered.}
\end{figure}

The consecutive interevent times of these users are plotted in fig 5
which helps us to visualize the dynamics of their activities. For
the blog user (see fig 5(a)), we observe the clustering of extremely
long interevent times which is also called mountain-valley-structure
found in many complex systems\cite{mountain1,mountain2}. For the
wiki user, fig 5(b) shows similar clustering but the interevent time
longer than one day are separated by many short intra-day interevent
times which are rare in blog-posting (compared with fig 5(a)). The
consequence is that the values of $Memory$ become rather small. The
definition of $Memory$ is as follow\cite{2}:
\begin{equation}
M_{k} = \frac{1} {{{n_\tau } - 1}}\sum\limits_{i = 1}^{{n_\tau } -
1} {\frac{{({\tau _i} - {m_1})({\tau _{i + k}} - {m_2})}} {{{\sigma
_1}{\sigma _2}}}},
\end{equation}where $\tau_i$ is the interevent time values and $n_\tau$ is the
number of interevent time and $m_1(m_2)$ and $\sigma _1(\sigma _1)$
are sample mean and sample standard deviation of $\tau_i$'s
($\tau_{i+k}$'s). The two interevent times $\tau_i$ and
$\tau_{i+k}$ are separated by $k$ events. The $Memory$ $M_1$) of the
blog user is 0.13 but the one of the wiki user is only 0.02.

The average $M_{k}$ of all qualified users with $k$ ranging from 1
to 35 is shown in fig 6. Average ${M_{1}}$ of wiki-revising is 0.13
which is obviously less than 0.21, the $M_1$ in blog-posting. This result
is in agreement with the one we found in Fig 5(a) and (b). As there
are different mechanisms in human activity in the intra- and
inter-day range, we find a way to study the memory of these
mechanisms separately. We remove the interevent times of
wiki-revising which is less than 1000 minutes (about 1 day) and
analyze the remaining series which only contain the inter-day
intervals. This allows us to consider only the memory in the
inter-day intervals and ignore the actions within one day. Figure
5(c) shows the interevent time series after data removal, of which
$M_{1}$ is 0.12. Correspondingly, we also find a significant
increase in the average $M_{1}$ of wiki-revising through this
procedure. As shown in the inset of fig 6, average
$M_1$ increases to 0.20 which is very close to the one
in blog-posing. Moreover, the decay curve is similar to that of
blog-posting: when $k<10$, it decays asymptotically as a power law;
when $k>10$, it decreased exponentially.

\section{Discussion}

We conclude by remarking two concrete evidences which support our
conjecture that human activity patterns are significantly different
in different time scale. Firstly, the exponents of interevent time
distribution is different in the intra- and inter-day range. In
addition to comparison with the previous empirical studies, we show
difference at the individual and global level by investigating the
activity patterns of wiki-revising and blog-posting. The second
evidence is the different dependence on $Activity$: for the
inter-day range, the exponents increase with $Activity$; for the
intra-day range, the exponents decrease with $Activity$ and in smaller
magnitude. On the other hand, we show the behavioral similarity
between wiki-revising and blog-posting as the same exponent
dependence is observed in corresponding range. This similarity
further increases after removal of the intra-day interevent times of
wiki-revising. Previous study reported the lack of memory in human
activity but our work shows that the presence of intra-day
activities mask the correlation between consecutive inter-day
activities and lead to an underestimate of memory. Can we thus
classify human activities by the interevent time scale? How to
accurately measure the memory in a series which is complex and
heavy-tailed? Further investigations are required in these directions.

In our previous studies\cite{dblog1}, the personal-preference model
was suggested to describe blog-posting, which successfully generate the
exponent dependence on $Activity$ and the significant memory. Here,
our analysis further shows that the model is suitable for
wiki-revising in the inter-day range as it shows the same exponent
dependence. However, there is still no model which can explain the
negative relationship between the exponents and $Activity$ in
intra-day range. One possible explanation is the time scale in
scheduling activities. We can plan our daily schedule carefully
according to our personal preference but we hardly plan what
to do every minutes. Our actions in minutes are more stochastic
which may lead to the smaller burstiness in the intra-day range (the
exponents in this range is often smaller). Though random walk in one
dimension \cite{rev1,dblog2} can be used to explain interevent time
in stochastic process, the value of exponent obtained is fixed to be
1.5 which does not agree with the present empirical result.

We finally remark again the interesting behaviors in both the intra-
and inter-day range. There are interesting details within both intra- and
inter-day range. A slight hump is observed in $P(\tau)$ at $\tau
\approx 1$ hours. For inter-day range, the decay of memory is
power-law when $k < 10$ and became exponential beyond this range. Is
there a relationship between time units (such as minute, hour, week,
month) and the dynamics underlying human activities? For example,
trend change observed in $P(\tau)$ at one hour may due to the timing
of tasks in hours.

\acknowledgments This work was funded by the National Basic Research
Program of China 973 Program No. 2006CB705500 , the National
Natural Science Foundation of China Grants No. 10975126 and No.
10635040 , the Specialized Research Fund for the Doctoral Program of
Higher Education of China Grant No. 20093402110032


\begin{thebibliography}{0}

\bibitem{dm7}
  \Name{A-L.Barab\'asi.}
  \REVIEW{Nature}{435}{2005}{207}.

\bibitem{dm8}
  \Name{A.V\'azquez., J.G. Oliveira., Z.Dezso., K-I.Goh., I.Kondor. \and A-L.Barab\'asi.}
  \REVIEW{Phys. Rev. E}{73}{2006}{036127}.

\bibitem{dm1}
  \Name{R.D.Malmgren., D.B.Stouffer., A.E.Motter. \and L.N.Amaral.}
  \REVIEW{Proc. Natl Acad. Sci. U.S.A.}{105}{2008}{18153}.

\bibitem{dm2}
  \Name{R.D.Malmgren., D.B.Stouffer., A.L.O.Campanharo. \and L.N.Amaral.}
  \REVIEW{Science}{325}{2009}{1696}.

\bibitem{dlett1}
  \Name{J.G.Oliveira. \and A-L.Barab\'asi.}
  \REVIEW{Nature}{437}{2005}{1251}.

\bibitem{dweb1}
  \Name{ B.Goncalves. \and J.J.Ramasco.}
  \REVIEW{Phys. Rev. E}{78}{2008}{026123}.

\bibitem{dweb2}
  \Name{F.Radicchi.}
  \REVIEW{Phys. Rev. E.}{80}{2009}{026118}.

\bibitem{dmessage1}
  \Name{W.Hong., X-P.Han., T.Zhou. \and B-H.Wang.}
  \REVIEW{Chinese. Phys. Lett.}{26}{2009}{028902}.

\bibitem{dmov1}
  \Name{T.Zhou., H.A.T.Kiet., B.J.Kim., B-H.Wang. \and P. Holme.}
  \REVIEW{EPL}{82}{2008}{28002}.

\bibitem{dgame1}
  \Name{A.Grabowski., N.Kruszewska. \and R.A.Kosi\'nski.}
  \REVIEW{Phys. Rev. E}{78}{2008}{066110}.

\bibitem{m2}
  \Name{X-P.Han., T.Zhou. \and B-H.Wang.}
  \REVIEW{New. J. Phys.}{10}{2008}{073010}.

\bibitem{m4}
  \Name{A.V\'azquez.}
  \REVIEW{Physica A}{373}{2007}{747}.

\bibitem{m3}
  \Name{J.G.Oliveira. \and A.V\'azquez.}
  \REVIEW{Physica A}{388}{2009}{187}.

\bibitem{dlinux1}
  \Name{S.K.Baek., T.Y.Kim. \and B.J.Kim.}
  \REVIEW{Physica A}{387}{2008}{3660}.

\bibitem{dco1}
  \Name{N.N.Li., N.Zhang. \and T.Zhou.}
  \REVIEW{Physica A}{387}{2008}{6391}.

\bibitem{dprint1}
  \Name{U.Harder. \and M.Paczuski.}
  \REVIEW{Physica A}{361}{2006}{329}.

\bibitem{dm7}
  \Name{A-L.Barab\'asi.}
  \REVIEW{Nature}{435}{2005}{207}.

\bibitem{dblog1}
  \Name{P.Wang., T.Zhou., X-P.Han. \and B-H.Wang.}
  arXiv:1007.4440v2.

\bibitem{2}
  \Name{K-I.Goh. \and A-L.Barab\'asi.}
  \REVIEW{EPL}{81}{2008}{48002}.

\bibitem{memory2}
  \Name{K.Yamasaki., L.Muchnik., S.Havlin., A.Bunde. \and H.E.stanley.}
  \REVIEW{Proc. Natl Acad. Sci. U.S.A.}{102}{2005}{9424}.

\bibitem{memory3}
  \Name{D.Rybski., S.V.Buldyrev., S.Havlin., F.Liljeros. \and H.A.Makse.}
  \REVIEW{Proc. Natl Acad. Sci. U.S.A.}{106}{2009}{12640}.

\bibitem{blog1}
  \Name{A.Grabowski.}
  \REVIEW{Eur. Phys. J. B}{69}{2009}{605}.

\bibitem{blog2}
  \Name{R.Kumar., J.Novak., P.Raghavan. \and A.Tomkins.}
  \REVIEW{In
Proceeding of the12th International Conference on World Wide
Web}{}{2003}{568}.

\bibitem{book}
  \Name{D-R.He., Z-H.Liu. \and B-H.Wang.}
  \Book{complex systems and complex
networks}
  \Publ{Higher Education Press of China, Bejing}
  \Year{2009}.

\bibitem{mountain1}
  \Name{D.Rybski. \and A.Bunde.}
  \REVIEW{Physica A}{388}{2009}{1687}.

\bibitem{mountain2}
  \Name{S.lennartz., V.N.Livina., A.Bunde. \and S.Havlin.}
  \REVIEW{EPL}{81}{2008}{69001}.

\bibitem{rev1}
  \Name{M.E.J.Newman.}
  \REVIEW{Contemporary Physics}{46}{2005}{323}.

\bibitem{dblog2}
  \Name{M.Gotz., J.Leskovec., M.McGlohon. \and C.Faloutsos.}
  \REVIEW{Proceedings of the Third International ICWSM Conference}{}{2009}{}.



\end{thebibliography}
\end{document}